\documentclass[doublespacing,preprint]{elsart}

\usepackage{epsfig}

\usepackage{amssymb}

\newcommand{\nn}{{ {n_+}}}

\newcommand{\NN}{{ {N_{run}}}}

\begin{document}

\begin{frontmatter}

\title{
Optimal 
Proton Trapping Strategy for
a Neutron Lifetime Experiment
}


\author{Kevin J. Coakley\thanksref{label2}\corauthref{cor1}}
\ead{kevin.coakley@nist.gov}

\corauth[cor1]{
Contributions of NIST staff to this work are not subject to copyright laws in the US.
}


\address[label2]{Statistical Engineering Division, National Institute of Standards and Technology, 325 Broadway,
Boulder C0 80305
fax: 303 497 3012
phone: 303 497 3895
}
\author{}

\address{}

\begin{abstract}
In a neutron lifetime experiment
conducted at
the National Institute of Standards and Technology,
protons produced by neutron decay
events are
confined in a proton trap.
In each run of the experiment, there is a trapping stage 
of
duration $\tau$.
After the trapping stage,
protons are purged from the trap.
A proton detector provides incomplete information because it goes dead after detecting
the first of any purged protons. 
Further,
there is a dead time $\delta$ between the
end of the trapping stage in one run and the beginning of the next
trapping stage in the next run.
Based on the fraction of runs where
a proton is detected, 
I estimate the trapping rate
$\lambda$ by the method of maximum likelihood.
I show that the expected value of the maximum likelihood
estimate is infinite.
To obtain a maximum likelihood estimate with
a finite expected value and a well-defined and finite variance,
I restrict attention to a subsample of
all realizations of the data. This subsample excludes
an exceedingly rare realization that yields an
infinite-valued estimate of $\lambda$.
I present asymptotically valid formulas
for the bias, root-mean-square prediction error,
and standard deviation of the maximum 
likelihood estimate of $\lambda$ for this subsample.
Based on nominal values of $\lambda$ and the dead time $\delta$,
I determine the optimal duration of the trapping
stage
$\tau$ by minimizing the root-mean-square prediction error of the
estimate.
\end{abstract}

\end{frontmatter}

\normalsize
\section{Introduction}

Ion traps play a key role in 
fundamental physics experiments (Ref. 1).
In this paper, I focus on
statistical methods for uncertainty analysis and
planning 
of
proton trap neutron lifetime experiments (Refs. 1-5)
and related experiments such as Ref. 6.
When a neutron decays, it produces a proton, an electron and an
antineutrino.  An accurate determination of the mean lifetime of the
neutron is critically important for testing the fundamental theories
of physics (Ref. 7). Further, the mean lifetime of the neutron is an important
parameter in the astrophysical theory of big bang nucleosynthesis (Ref. 8).
In a proton trap neutron lifetime experiment performed at
the National Institute of Standards and Technology (NIST),
a beam of neutrons passes through a detection
volume.  Based on measurements of the neutron flux and the proton production
rate, one measures the mean lifetime of the neutron.  Each run of
the experiment consists of trapping stage where protons are confined
in a trap (Refs. 2-5), and a detection stage.
The detector provides incomplete information because it goes dead after detecting
the first proton.
Based on the number of runs where
a proton is detected,
one can estimate the proton trapping rate.

In earlier work
[Ref. 9],
this estimation problem 
was studied using
a Bayesian method.
Given a particular realization of the
data
(the number of runs
where at least one ion (proton in this paper)
is trapped),
a formula for the
posterior mean
of the
ion trapping rate was
presented based on a prior probability model for the trapping rate.
In this work, I estimate the trapping rate
by the method
of
maximum likelihood 
and focus on the statistical properties of
this estimate.
I neglect physical sources of systematic error due to
effects such
as a time varying proton trapping rate or 
fluctuations in the actual trapping stage interval
about the nominal value sought by the experimenter.

In Section 2, 
I demonstrate that
the bias (expected value minus true value) and variance of the maximum likelihood
estimate of the trapping rate $\lambda$ are
infinite.
This is so because a rare realization of the data
yields
an infinite estimate of $\lambda$.
This technical problem can be dealt with in various
ways.
One  could
quantify uncertainty by constructing
confidence intervals of
finite width
even
though the variance of the estimate is infinite.
Another approach would be to
introduce a stopping rule 
so that the experiment is continued until
no protons are trapped in at least one run.
I do not pursue either of these approaches here.
Instead,
I restrict the sample space
to include only realizations of data where
one observes at least one run
where no protons are trapped.
For realizations of data in this subsample, the 
maximum likelihood estimate has finite first and second moments.
In Section 3, 
I derive
asymptotically valid formulas
for the bias, variance, and mean-square-error of a maximum
likelihood estimate of the proton trapping rate computed from this
subsample.
In general, one expects 
estimates that are nonlinear functions
of the observed data, such as the maximum likelihood
estimate of the trapping rate,
to be biased
[Ref. 10].
In Section 4,
where,
based on nominal values of trapping rate and dead time,
I determine the trapping time that
minimizes the root-mean-square prediction error of 
the maximum likelihood estimate of 
$\lambda$ in the subsample of interest.

\section{Statistical Model}
In a simulated proton trapping experiment there are many runs.
During each run,
I assume that
the duration of the proton trapping stage 
$\tau$ is an adjustable constant that is known
with negligible uncertainty.
During the trapping stage, I assume that protons are
trapped at a constant rate $\lambda $.
Further, I restrict attention to the case where
$\lambda > $ 0.
After the trapping stage,
protons are purged from the trap.
A proton detector provides incomplete information because it goes dead after detecting
the first of any purged protons. 
Further,
there is a fixed dead time $\delta$ between the
end of the trapping stage in one run and the beginning of the next
trapping stage in the next run.
I assume that $\delta$ is known with negligible uncertainty.
If the total time of the experiment is $T$, the total number of runs is 
\begin{eqnarray}
\NN =  INT( ~  \frac{T} { \tau + \delta} ~ ).
\end{eqnarray}
Above, the function $INT(x)$  rounds 
the continuous variable $x$ down to the nearest integer. 
Let
$\nn$ be the observed number of runs where at least one proton
is trapped.
I model 
the number of protons trapped during any run as
a realization of a Poisson process with expected value
$\lambda \tau$. Hence, the probability that 
no ion is trapped for a given run is 
\begin{eqnarray}
p_0 =  \exp( -\lambda \tau ).
\end{eqnarray}
The maximum likelihood estimate of
$p_0$ is 
\begin{eqnarray}
\hat{p}_o = 1 - \frac{\nn}{\NN} 
\end{eqnarray}
where $\nn$ is the number of runs where at least
one proton is trapped.
Thus,
the maximum likelihood estimate of
$\lambda$ is
\begin{eqnarray}
\hat{\lambda} =
 - \frac{1}{\tau} \ln  \hat{p}_0  =
 - \frac{1}{\tau} \ln ( 1 - \frac{\nn}{\NN} ) .
\end{eqnarray}

Since $\nn$ is a binomial random variable,
the probability that $\nn = k$ is $P(k)$, where
\begin{eqnarray}
P(k) = \frac{\NN!}{(\NN-k)!k!} (1-p_0) ^ k p_0 ^ {\NN-k} .
\end{eqnarray}
Hence, the expected value of the maximum likelihood estimate of $\lambda $
is
\begin{eqnarray}
E( \hat{\lambda} ) = 
 - \frac{1}{\tau}
\sum_{k=0}^{\NN} 
P(k)
\ln ( 1 - \frac{k}{\NN} ).
\end{eqnarray}
Similarly,
the expected squared value of the estimate 
is
\begin{eqnarray}
E( \hat{\lambda} ^2 ) = 
\frac{1}{\tau^2}
\sum_{k=0}^{\NN} 
P(k)
(\ln ( 1 - \frac{k}{\NN} ))^2.
\end{eqnarray}
For $\lambda > 0$,
$P( \NN ) = (1-p_0)^\NN  >  0$,
and 
both the expected value (first moment) and expected squared value
(second moment)
of 
$\hat{\lambda}$ 
are infinite.
The 
variance of $\hat{\lambda}$, $VAR( \hat{\lambda} )$,
is not defined
because
\begin{eqnarray}
VAR( \hat{\lambda} ) = 
E( \hat{\lambda}^2 ) -
(E(\hat{\lambda}))  ^2,
\end{eqnarray}
and 
both terms on the right hand side of
Eq. 8 are infinite.

To ensure that
both
$E( \hat{\lambda}  )$
and
$E( \hat{\lambda}^2  )$
are finite,
I restrict the
sample space to
realizations of the data where
$\nn< \NN$. 
From a practical point of view, this means that 
realizations of data where $\nn = \NN $ would be ignored.
For neutron lifetime experiments of current interest,
the probability that $\nn = \NN$ is negligible
provided that
$\tau$ 
is judiciously chosen.
Hence,
this subsampling restriction does not
significantly
affect data
collection procedures for neutron lifetime experiments
of current interest.
In this subsample, the discrete probability 
density
function for allowed realizations of $\nn = 0,1, \cdots, \NN - 1$ is
$P_{*}(k)$, where 
\begin{eqnarray}
P_{*}(k) = \frac{P(k)}{ 1 - P(\NN) } .
\end{eqnarray}
For this subsample, the first two moments of the maximum likelihood
estimate are
\begin{eqnarray}
E( \hat{\lambda} ) = 
 - \frac{1}{\tau}
\sum_{k=0}^{\NN-1} 
P_*(k)
\ln ( 1 - \frac{k}{\NN} ),
\end{eqnarray}
and
\begin{eqnarray}
E( \hat{\lambda} ^2 ) = 
\frac{1}{\tau^2}
\sum_{k=0}^{\NN-1} 
P_*(k)
(\ln ( 1 - \frac{k}{\NN} ))^2.
\end{eqnarray}
Since the first two moments (Eqns. 10 and 11) of $\hat{\lambda}$
are finite, the variance of
$\hat{\lambda}$
is defined and finite.
Next, I present analytical formulas
to approximate the fractional bias, fractional standard deviation and
fractional root-mean-square prediction error of the estimate computed for this
subsample.

\section{\label{sec:sec3}Asymptotic Approximations for Analysis in Subsample}

In the subsample where $n_+ < \NN$,
I derive asymptotically valid approximations for
the fractional bias ($FBIAS$),
fractional root-mean-square prediction error ($FRMS$),
and fraction standard deviation ($FSE$) 
of $\hat{\lambda}$ where
\begin{eqnarray}
FBIAS = \frac {
E(\hat{\lambda} - \lambda ) }
{\lambda},
\end{eqnarray}
\begin{eqnarray}
FRMS = 
\frac { \sqrt{ E( \hat{\lambda} - \lambda )^2 }  } 
{ \lambda },
\end{eqnarray}
and
\begin{eqnarray}
FSE =
\frac { \sqrt{ E( \hat{\lambda} - E( \hat{\lambda} ))^2 }  } 
{ \lambda } =
\sqrt{ (FRMS) ^ 2 - (FBIAS)^2 }.
\end{eqnarray}

To facilitate analysis of
$\hat{\lambda}$ in the subsample
I write 
\begin{eqnarray}
\hat{p}_0 =
p_0  - \epsilon
= p_0 ( 1  - \frac{\epsilon}{p_0} ),
\end{eqnarray}
\begin{eqnarray}
\epsilon = 
p_0 - \hat{p}_0.
\end{eqnarray}
Thus,
\begin{eqnarray}
\ln\hat{p}_0 = \ln( p_0 )  + \ln ( 1 - w )
\end{eqnarray}
where
\begin{eqnarray}
w = \frac{\epsilon}{p_0}.  
\end{eqnarray}
In the subsample, $w$ takes discrete values in the following interval
\begin{eqnarray}
1 - \frac{1}{p_0} \le w \le 
1 - \frac{1}{N_{run}p_0}.
\end{eqnarray}

The $BIAS$ of the maximum likelihood estimate of $\lambda$ in the
subsample is
\begin{eqnarray}
BIAS
= 
-{\tau}^{-1} 
(
E ( \ln\hat{p}_0 ) -  
\ln( p_0 )  
) 
=
-{\tau}^{-1} 
E (\ln ( 1 - w ) )
\end{eqnarray}
where $w$ is a random variable.

I derive an asymptotically valid
expression for $BIAS$
based on a
local
approximation
for $\ln( 1 - w)$
in the vicinity
of 0.
I approximate $f(w) = -\ln( 1 - w)$ as a fourth order polynomial $\hat{f}(w)$ 
where
\begin{eqnarray}
\hat{f}(w) \approx f(0) + w f^{1}(0) + \frac{w^2}{2!} f^2(0) + \frac{w^3}{3!}f^3(0)
+ \frac{w^4}{4!} f^4(0)
\end{eqnarray}
where $f^{k}(0)$ is the $k$th derivative 
of $f(w)$ evaluated at $w=0$.
Thus, 
\begin{eqnarray}
\hat{f}(w)  =
\sum_{k=1}^{4}
\frac{w^k}
{k}  = 
\sum_{k=1}^{4}
\frac{\epsilon^k}
{k (p_0)^k}.  
\end{eqnarray}

Since
\begin{eqnarray}
\epsilon = \frac{\nn - E(\nn)} {\NN},
\end{eqnarray}
the central moments 
$\mu_r = E(  ( \nn - E(\nn) )  ^ r ) $ 
are relevant.
In the full sample where $\nn \le \NN$,
$\nn$ is a binomial random variable with
an expected value equal to $\NN (1 - p_0)$.
Hence, in full sample, its first four central moments are
[11]
\begin{eqnarray}
\mu_1 = 0,
\end{eqnarray}
\begin{eqnarray}
\mu_2 = \NN p_0 ( 1 - p_0),
\end{eqnarray}
\begin{eqnarray}
\mu_3 = \NN p_0 ( 1 - p_0) (2 p_0 -1),
\end{eqnarray}
and
\begin{eqnarray}
\mu_4 = 3 N^2_{run} p^2_0 {(1-p_0)}^2 ~+~  \NN p_0 ( 1 - p_0) ( 1 - 6 p_0 (1-p_0 )).
\end{eqnarray}
Since
$P(\nn = \NN) $ tends to 0 exponentially as a function of $\NN$,
the asymptotic central moments of $\nn$ in the subsample
are given by
Eqns. 24-27. 

Based on Eqns. 22,25,26 and 27, I get the following approximation for
$FBIAS$ 
\begin{eqnarray}
\widehat{FBIAS}  \approx
\frac{1}{\lambda \tau}
[~\frac{\mu_2}{2 (\NN p_0)^2} 
+
\frac{\mu_3}{3 (\NN p_0)^3}  
+\frac{\mu_4}{4 (\NN p_0)^4}
~].
\end{eqnarray}

In a similar calculation where I approximate 
$f(w)$ as a second order polynomial,
I get the following approximation for
$FRMS$
\begin{eqnarray}
\widehat{FRMS}  
\approx
\frac{1}{\lambda \tau}
\sqrt{
\frac{\mu_2}{(\NN p_0)^2} 
+\frac{\mu_3}{(\NN p_0)^3}  
+\frac{\mu_4}{4 (\NN p_0)^4}
}.
\end{eqnarray}
From Eqns. 28 and 29,
I derive an approximation for 
$FSE$ 
\begin{eqnarray}
\widehat{FSE} = \sqrt{ (\widehat{FRMS}) ^ 2 - (\widehat{FBIAS})^2 }.
\end{eqnarray}

Since
the asymptotic standard deviation of the random variable $w$
is
\begin{eqnarray}
\sigma_w = \sqrt{ \frac{1-p_0}{\NN p_0}},
\end{eqnarray}
I expect Eqns. 28-30 to be
asymptotically valid as
$\NN$ increases to large values.
Next,
I
present evidence
consistent
with this expectation
for a number of cases
(Tables 1,2).

\section{Example}
For particular cases, 
I compute the actual values of
$FRMS$, $FBIAS$ and $FSE$ 
using Eqns. 9-14.
I set
$\delta =  $ 100 $\mu $s (0.0001 s)
and
$\lambda = $ 1 Hz
because these are typical values
for 
experiments done at NIST.
For experiments of total duration of $T=$ 10 s ,25 s, 50 s, 100 s, 200 s, and 400 s,
$FBIAS$ was much less than $FRMS$ (see Figure 1 and Table 1).
Furthermore,
$FBIAS$ was more sensitive to
$\tau$ than $FRMS$ was (Figure 1).
For the cases summarized in Table 1,
the fractional systematic error and fractional RMS prediction error
are well approximated as
$FBIAS \propto T^{-1}$,
and
$FRMS \propto T^{-1/2}$.
For each cases,
both
$FRMS$ and $\widehat{FRMS}$ (Eqn. 29)
took their minimum values at
$\tau = $ 0.014 s. 
The resolution of the grid on which I computed
RMS prediction errors is 0.001 s in the neighborhood of 
the 0.014 s.

In a second simulation,
I set
$T = $ 400 s and $\lambda = $ 100 Hz.
I vary $\tau$ so that the expected number of
trapped protons per run, $\lambda \tau$, varies from .001 to 4.
For these cases, 
$\widehat{FBIAS}$
and
$\widehat{FRMS}$
closely track the actual values of
$FBIAS$ and $FRMS$ (Table 2).
For the smallest values of $\tau$,
the accuracies of the approximations
are highest. I attribute the slight degradation
of approximation accuracy
at
the largest values of $\lambda \tau$  to the
fact 
that
$\sigma_w$ (Eq. 31) is a monotonically increasing function
of
$\tau$.

For convenience,
I express the  fractional RMS prediction error
of $\hat{\lambda} $ as
\begin{eqnarray}
FRMS
= 0.001\sqrt{ \frac{ T^* }{T} },
\end{eqnarray}
where $T$ is the total time of the experiment.
The parameter $T^{*}$ is a function of
$\lambda$, $\tau$ and $\delta$.
In proton trap neutron lifetime experiments,
$\lambda$
depends on experimental 
details
including the length of the trap;
trapping efficiency; and
the neutron flux
(Refs. 2-5)
For the cases considered here,
I compute $T^{*}$ directly using Eqns. 9-14 and Eq. 32.
For fixed values of $\lambda$, $\delta$,
and $\tau$,
the derived value of
$T^*$
is approximately the same for all values of
$T$.
An exception to this rule is
when $\delta$ is large
and there are very few bins. I attribute this to truncation effects
associated with rounding $\NN$ to an integer (Eq. 1).
Thus,
one can compute $T^{*}$ 
from simulation data corresponding to one sufficiently large value of $T$
and predict $FRMS$ at other large values of $T$. 
As a caveat, for very short experiments, the asymptotic
theory may not apply and a direct simulation may be
necessary.

For the case where the dead time $\delta$ is fixed,
$T^{*}$  varies as a function of both
$\lambda$ and $\tau$ (Figure 2) in a complicated
manner.
To
clarify resutls,
I scale
$T^{*}$ and $\tau$ by 
the true trapping rate $\lambda$ (Figure 3).
I define $\tau_{opt}$ to be the value of $\tau$ that minimizes
$FRMS$.
Based on Figure 3,
the most elucidating way to
find the optimal
data collection strategy is
to minimize $\lambda T^{*}$ as a function of
$\lambda \tau$.
For the cases shown in Figures 2 and 3,
I conclude that $\lambda \tau_{opt}$
increases 
as $\lambda$ increases.
In a second simulation experiment, I consider
cases where
$\lambda$ is fixed but the dead time $\delta$ 
varies from case to case.
For these cases,
as $\delta$ increases,
so too does $\lambda \tau_{opt}$ (Figure 4).

\section{Discussion}

Earlier I stated that
the subsample restriction 
has no practical effect on data collection
for neutron lifetime experiments of current interest.
To make this claim more concrete,
I compute 
the probability of observing $\nn = \NN$ in the full sample
for the cases listed in Table 1.
For these cases,
$\tau = $ 0.014 s
and
$\lambda = $ 1 s${}^{-1}$
and
$P(\NN) \approx$ 
$10 ^ { -1.8569 \NN} $.
Hence for an experiment of total duration 100 s,
$P(\NN) \approx 
10 ^{-13169}$.

In the study, I quantified
$FBIAS$ given knowledge of $\lambda$
and
particular values of
$\tau$, $\delta$ and $T$.
In actual experiments, one would
use the estimated 
value of $\lambda$ rather than the true value.
Hence, in Eqn. 28, one would use 
$\hat{p}_0 = \exp( -\hat{\lambda}\tau)$ rather
than the true value of $p_0$.
If $FBIAS$ is negligible,
there is no need to correct $\hat{\lambda}$ for bias.
In principle,
when bias is significant,
a bias-corrected maximum likelihood estimate 
should be obtained using the following 
iterative procedure:
\begin{eqnarray}
\hat{\lambda}_{(k+1,BC)} = 
\frac { \hat{\lambda}  }
{ 1 + \widehat{FBIAS}(\hat{\lambda}_{(k,BC)},\NN,\tau)
}
\end{eqnarray}
where
$\hat{\lambda}_{(0,BC)} = \hat{\lambda}$
and
$\hat{\lambda}_{(k+1,BC)}$ is the 
bias-corrected maximum likelihood estimate at
the $k$th iteration.
In practice, one iteration of the above procedure
may yield a numerically stable estimate of the bias-corrected
maximum likelihood estimate for cases of interest.

\section{Summary}
In this work,
I studied the statistical properties of
a maximum likelihood estimate of
the rate at which protons are trapped.
This study is relevant to
proton trap neutron lifetime experiments
at NIST and similar experiments elsewhere.
After the first proton is detected, the detector
goes dead.
Hence, the detector provides incomplete information.
Due to this incompleteness, I showed that
the first two moments of the
maximum likelihood estimate of the trapping rate $\lambda$ 
are infinite. Hence, the variance of the maximum likelihood
estimate is not defined.
To construct a maximum likelihood estimate with a finite variance,
I restricted attention to a
subsample of realizations of the data
that excludes an exceedingly rare realization of the data that yields an
infinite valued estimate  of $\lambda$. 
I demonstrated that the probability of observing
this rare realization
quickly decreases to a negligible
value
for
a 
judicious 
choice of the trapping time
for proton trapping rates
achievable at NIST  (Secion 5).
Hence, restricting attention to 
the subsample of interest has no
practical effect on current neutron lifetime
experiments of interest.
Based on the discrete probability
density function for this subsample,
I derived exact formulas for the first two moments of the
maximum likelihood estimate of $\lambda$ (Eqns. 10 and 11).
I derived asymptotically valid formulas
for the fractional bias, fractional RMS prediction
error and
fractional standard deviation of the
maximum likelihood estimate 
(Eqns. 28-30).
I showed that
the approximation error associated with
these
formulas is
low for a variety of cases
(Tables 1,2).

I demonstrated
that
the fractional bias ($FBIAS$) of the
estimate
was more sensitive to $\tau$ than the fractional mean-square prediction error
($FRMS$)
was (Figure 1).
As as a function of total
observing time $T$, I showed that 
$FBIAS$
decreases 
much faster than does
$FRMS$
(Table 1).

I presented an objective method to select the optimal value of $\tau$
by minimizing $FRMS$.
In general, the optimal 
trapping time $\tau$
that minimizes $FRMS$
is a complicated function of both 
dead time $\delta$ and
the trapping rate $\lambda$ (Figures 2-4).
For experimental planning purposes,
my asymptotic
approximations
(Eqns. 28-30)
should be useful for determining
the optimal data collection strategy
and for 
quantifying random and systematic errors.

In this study, I neglected physical sources of systematic error due to
effects such
as a time varying proton trapping rate or 
fluctuations in the actual trapping stage interval
about the nominal value sought by the experimenter.
Hence, the bias I quantified here
is
a purely statistical artifact due to the fact
the maximum likelihood estimate of the trapping rate
is a nonlinear function of the observed data.

\newpage{}
\begin{table}
Table 1.
Proton trapping rate is
$ \lambda $ = 1 s${}^{-1}$. Dead time is $\delta = $ 0.0001 s.
Trapping stage duration is
$\tau = $ 0.014 s.
$FRMS$ and $FBIAS$ are
fractional
bias and fractional root-mean-square prediction error
of the maximum likelihood estimate of $\lambda$.
The approximations for 
$\widehat{FRMS}$ and $\widehat{FBIAS}$ are computed using Eqns.
28 and 29.
\begin{center}
\begin{tabular}{cccccccccc} \hline\hline
\\
\hline
$T$ (s) & $\NN$ & $E(\nn)$ & $FBIAS$ & $\widehat{ FBIAS}$ & $FRMS$ & $\widehat{FRMS}$ 
\\
10& 709& 9.85684& 0.000710859& 0.000710858& 0.318749& 0.318742
\\
25& 1773& 24.6491& 0.0002841& 0.0002841& 0.201479& 0.201477 
\\
50&  3546& 49.2981& 0.000142023& 0.000142023& 0.142446& 0.142446 
\\
100& 7092& 98.5962& 7.10046e-05& 7.10046e-05& 0.100717& 0.100717 
\\
200& 14184& 197.192& 3.55006e-05& 3.55006e-05& 0.0712154& 0.0712154 
\\
400&  28368& 394.385& 1.77499e-05& 1.77499e-05& 0.050356& 0.050356 
\\
\hline
\end{tabular}
\end{center}
\end{table}   
\newpage{}
\begin{table}
Table 2.
Simulation study.
Proton trapping rate is
$ \lambda $ = 100 s${}^{-1}$. Dead time is $\delta = $ 0.0001 s.
$T=$ 400 s.
\begin{center}
\begin{tabular}{cccccccc} \hline\hline
\\
\hline
$\tau $ (s)&$ \lambda \tau$ & $\NN$ & $E(\nn)$ & $FBIAS$ & $\widehat{ FBIAS}$ & $FRMS$ & $\widehat{FRMS}$ 
\\
1e-05&0.001& 3636360& 3634.55& 1.37569e-07& 1.37569e-07& 0.0165873& 0.0165873 
\\
1e-04&0.01& 2e+06& 19900.3& 2.51254e-07& 2.51254e-07& 0.00708878& 0.00708878 
\\
0.001&0.1& 363636& 34604.5& 1.4461e-06& 1.4461e-06& 0.00537793& 0.00537793 
\\
0.01&1&39603& 25033.9& 2.1695e-05& 2.1695e-05& 0.00658726& 0.00658698 
\\
0.015&1.5& 26490& 20579.3& 4.38173e-05& 4.38173e-05& 0.007644& 0.007643 
\\
0.02 &2.0&19900 &17206.8& 8.02887e-05& 8.02887e-05& 0.00896179& 0.00895891
\\
0.03 &3.0&13289 &12627.4& 0.000239664& 0.000239664& 0.0126488& 0.0126306 
\\
0.04 &4.0&9975& 9792.3& 0.000674739& 0.000674717& 0.0184136& 0.0183143
\end{tabular}
\end{center}
\end{table}

\begin{figure}
\vspace{0.01in}
\centerline{\epsfysize=5.9in
\epsffile{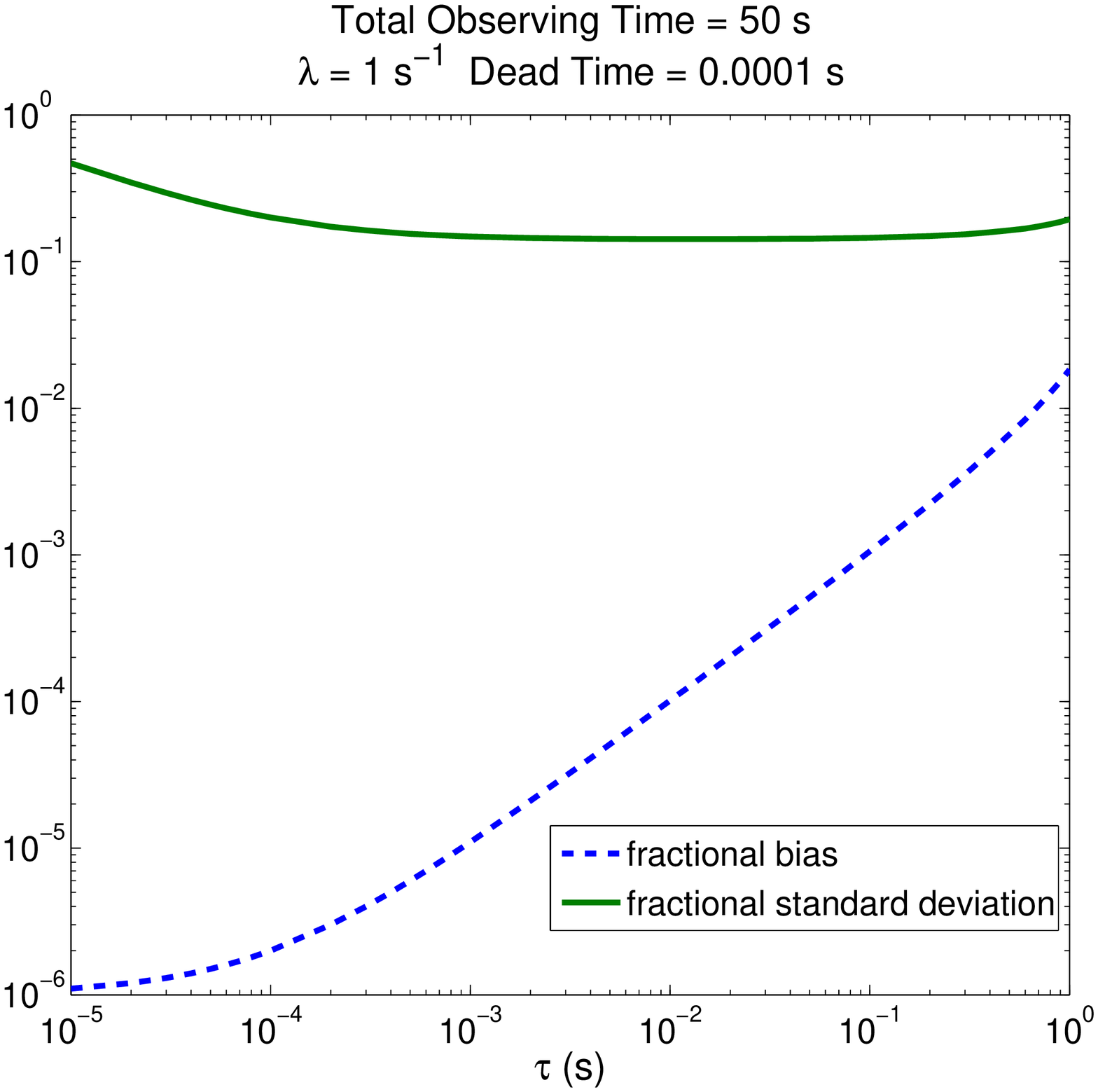}}
\vspace*{-0.05in}
\noindent
Figure 1.
Fractional bias ($FBIAS$) and fractional standard error ($FSE$) for simulation experiment where
true trapping rate is $\lambda = $ 1 s$^{-1}$,
total duration of experiment is $T=$ 50 s,
dead time is $\delta = $ 100 $\mu$s,
and trapping stage duration $\tau$ varies. 
\end{figure}

\begin{figure}
\vspace{0.01in}
\centerline{\epsfysize=5.9in
\epsffile{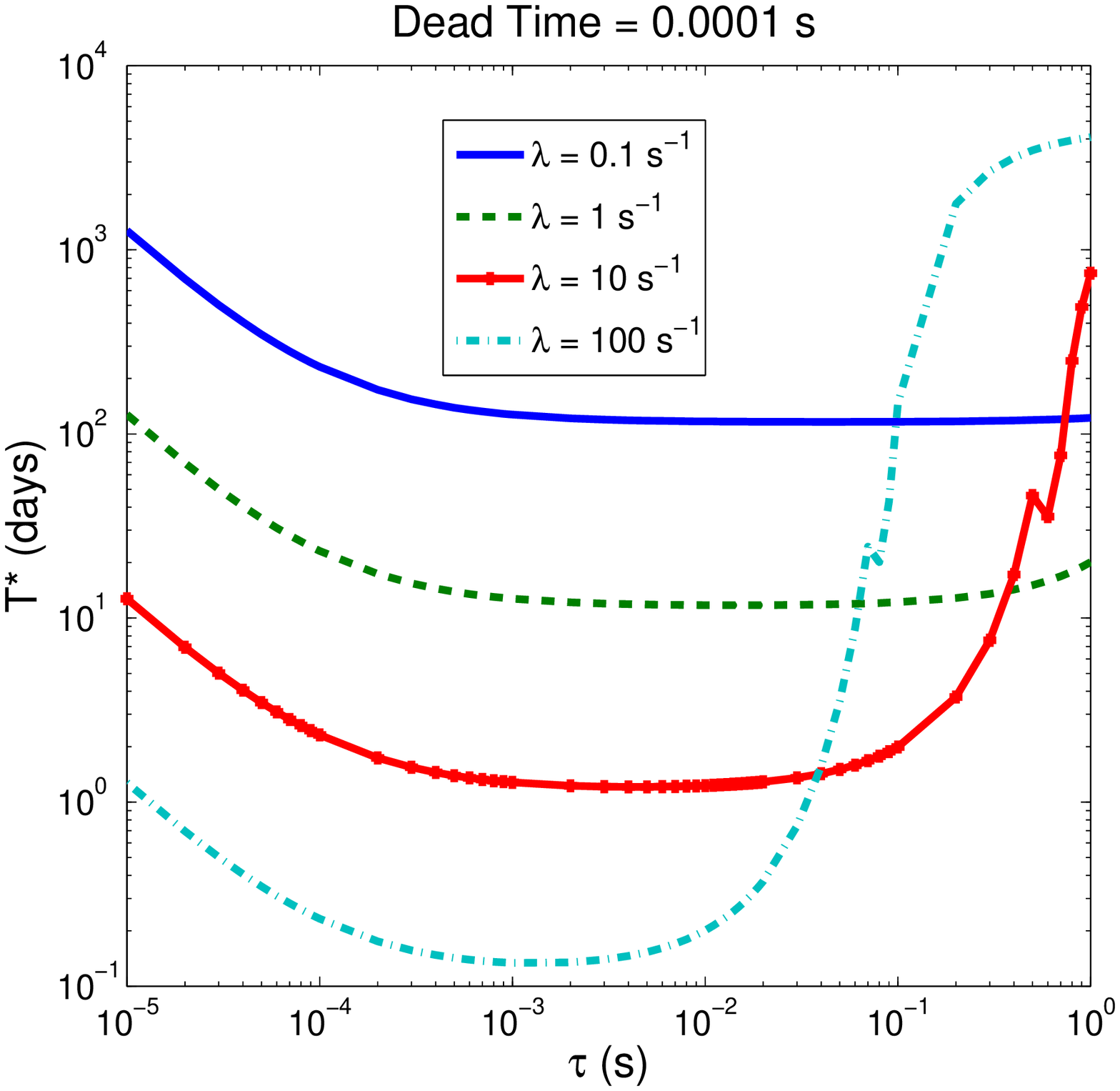}}
\vspace*{-0.05in}
\noindent
Figure 2.
Fractional RMS prediction error ($FRMS$) of the trapping rate is
expressed as
$0.001\sqrt{ \frac{ T^* }{T} }$,
where the total length of the experiment is $T$.
Dead time is $\delta = $ 0.0001 s.
\end{figure}

\begin{figure}
\vspace{0.01in}
\centerline{\epsfysize=5.9in
\epsffile{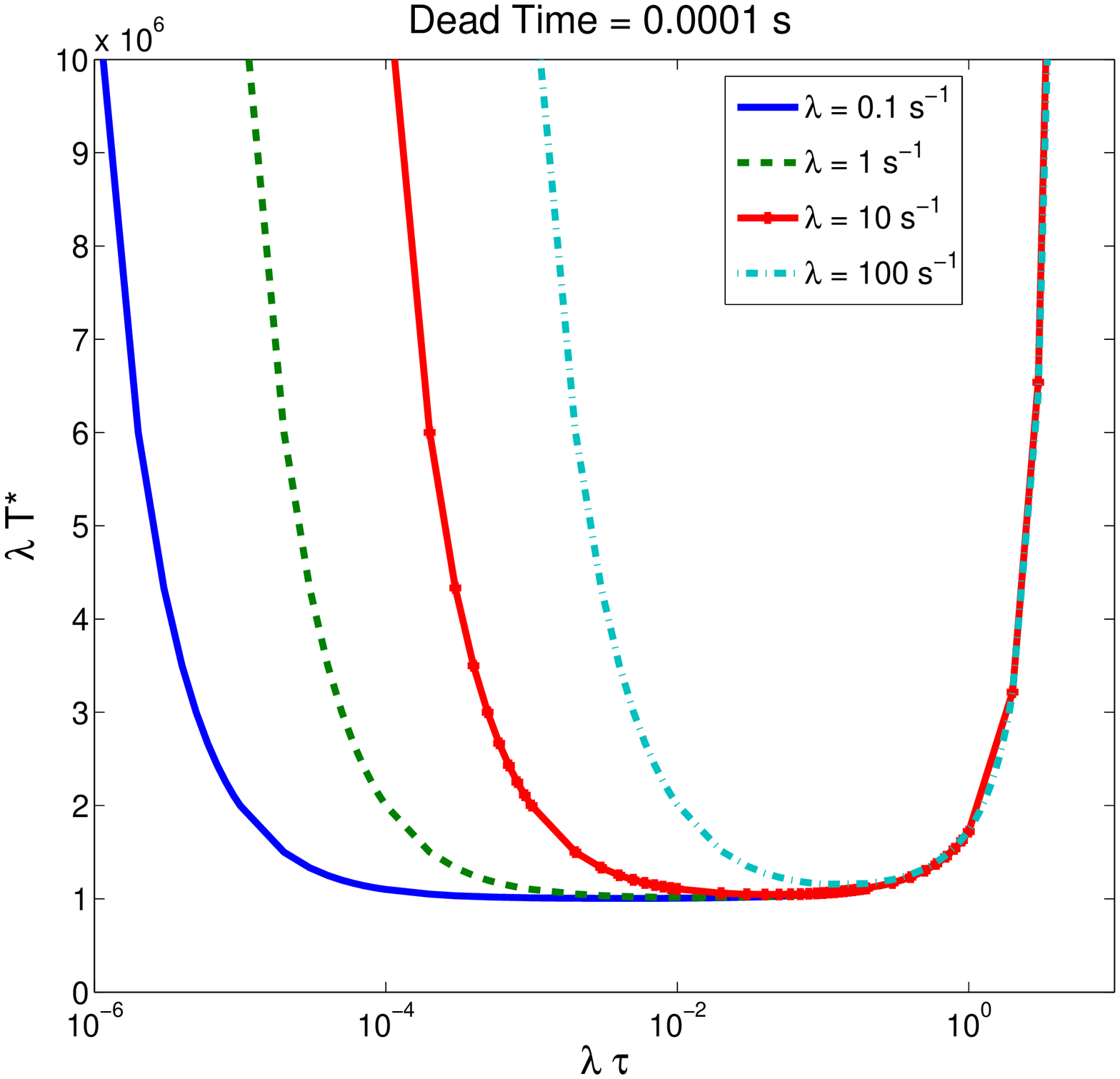}}
\vspace*{-0.05in}
\noindent
Figure 3.
Same results as in Figure 2,
but
$T^{*}$
and $\tau$ are 
rescaled to clarify results.
\end{figure}

\begin{figure}
\vspace{0.01in}
\centerline{\epsfysize=5.9in
\epsffile{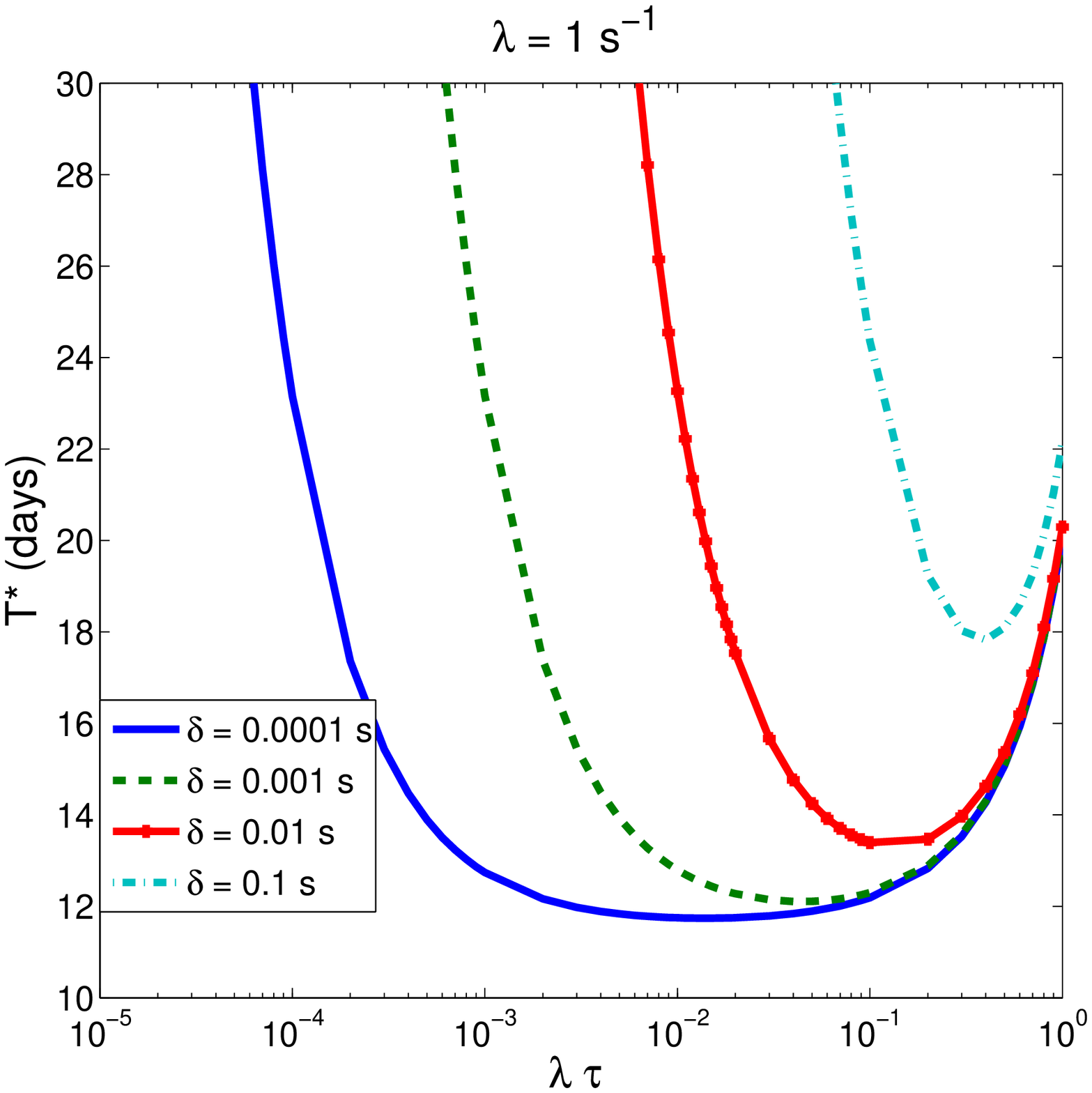}}
\vspace*{-0.05in}
\noindent
Figure 4.
True proton trapping rate is $\lambda = $  1 s${}^{-1}$.
\end{figure}

\end{document}